\documentclass[prd,aps,floats,twocolumn,nofootinbib]{revtex4}
\pdfoutput=1
\usepackage{latexsym}
\usepackage{amssymb}
\usepackage{color}
\usepackage{graphicx}

\begin{document}

\title{Realistic Cyclic Magnetic Universe}
\author{L. G. Medeiros}
\email{leogmedeiros@gmail.com}
\affiliation{Escola de Ci\^{e}ncia e Tecnologia, Universidade Federal do Rio Grande do
Norte. Campus Universit\'{a}rio, s/n, CEP 59072-970, Natal, Brazil}

\begin{abstract}
This work presents a complete cyclic cosmological scenario based on
nonlinear magnetic field. It is constructed a model composed by five fluids
namely baryonic matter, dark matter, radiation, neutrinos and a cosmological
magnetic field. The first four fluids are treated in the standard way and
the fifth fluid, the magnetic field, is described by a nonlinear
electrodynamics. The free parameters are fitted by observational data (SNIa,
CMB, extragalactic magnetic fields, etc) and by simple theoretical
considerations. As result arises a cyclic cosmological model which preserves
the main successes of standard big bang model and solve some other problems
like the initial singularity, the present acceleration and the Big Rip.
\end{abstract}

\bigskip

\vskip2pc
 \maketitle

\section{Introduction}

Although the inflationary scenario still maintains its status of the
paradigm in Cosmology, there is an increasing interest in the
alternative proposal of models displaying a bounce
\cite{pbb,ekpy,Boun,leo,Herman}. This means the possible existence of a
collapsing era prior to the actual expanding phase. One of the main
differences of these two proposals concerns the behavior of small
perturbations and its evolution from a perfect spatially homogeneous
phase to the production of inhomogeneities. There are hopes, due to
a through analysis of these processes \cite{nelson}, that this
question will be settled in the near future. Another important point
in favor of bouncing models is that they avoid the problem of
initial singularity. The present paper examines one of these
scenarios which the dynamical history of the universe is partial
controlled by a magnetic field. It is analyzed its main effects and its
compatibility with actual observations. The main hypothesis of this
model concerns the non-linearity character of magnetic field.

The effects of a nonlinear electromagnetic theory in a cosmological
setting have been studied in several articles
\cite{novello1,novello2,port1,port2,port3,port4,port5,port6} and for
different reasons. For example, in \cite{novello1} the non-linearity
is responsible to avoid the initial singularity, and in
\cite{novello2} the non-linearity is accountable for generate the
recent acceleration. In these two papers the framework is the same:
a cosmological magnetic field governed by a nonlinear
electromagnetic theory is responsible by the desirable effects.

Recently, these two effects were combined generating a cyclic
cosmological toy model \cite {Aline}. The purpose of the present
paper is improve this model and investigate how realistic it can be.
It is a well-known fact that the standard cosmological model
unavoidably leads to a singular behavior of the curvature invariants
in what has been termed the Big Bang. On the other hand, there are
some evidences that point in the direction that the universe is
undergoing an accelerated expansion \cite{Riess1998,Perm}. In
principle, these two problems have no simple connection and thus
does not have a unique combined solution. It should be tempted to
try to unify these two questions in a single model. It will be shown
that in the framework of a magnetic universe these two problems are
solved at once.

In general, the cyclic cosmological models can be divided in two classes:
Those which generate the cyclic behavior with non-conventional matter
fields \cite{Quintoncyclic,cyclic1}, and those which produce the
cyclicity through of extension of General Relativity
\cite{QuintonExt1,QuintonExt2,QuintonExt3}. The model proposed here belongs
to the first class where the non-conventional matter is represented by
a non-linear magnetic field.

The plan of the paper is as follows. Section II presents the notion
of the Magnetic Universe and its generic features. In section III it
is fitted or constrained the five independent parameters presents in
this models. The section IV presents the complete scenario
consisting of the five eras. The paper ends in section V with some
generic final comments. Moreover, it is presented in the appendix A
a brief review about the average procedure and in appendix B a
discussion about the model stability.

\section{Magnetic universe} \label{Sec-MU}

To construct a realistic magnetic universe we start assuming that the
universe is composed by the five fluids namely baryonic matter, cold dark
matter, neutrinos, radiation and a cosmological magnetic field. The first
four fluids will be treated in the standard way. Thus baryonic and dark
matter are modeling by non-relativistic fluids (zero pressure), and
neutrinos and radiation are featured as ultra-relativistic fluid (The mass
of neutrinos is neglected). The fifth fluid, the magnetic field, will be
described by the Lagrangian for the non-linear electrodynamics given by

\begin{equation}
L_{NLED} = \alpha^{2} \, F^{2} -\frac{1}{4} \, F - \frac{\mu^{2}}{F} + \frac{%
\beta^{2}}{F^{2}},  \label{lag}
\end{equation}
where the dimensional constants $\alpha, \beta$ and $\mu$ are to be
determined by observation.

The nonlinear terms can be interpreted as a phenomenological
approach. But a more interesting and fundamental scenario is suppose
these terms represent a classical interpretation of vacuum
polarization. Indeed, Heseinberg, Schwinger and others showed that
quantum corrections due vacuum polarization changes the classical
Lagrangian introducing nonlinear terms \cite{Euler,Schwinger}.

Let us suppose that the five fluids are independent. Moreover, the
magnetic field enter in cosmological scenario through the average
procedure described in appendix A. By construction, the average of
the electric part of the cosmological electromagnetic field must
vanishes, i.e. $E^{2} =0$. And in this case, each term of $L_{NLED}$
becomes cosmological independent in the sense that each one behaves
as a non-interacting perfect fluid (for details see \cite{Aline}).

Treating baryonic matter and dark matter as a single fluid and neutrinos and
radiation as another single fluid, the magnetic universe is completely
featured for six independent components. In this context, the total energy
density and the total pressure is given by
\begin{equation}
\rho_{T} = \rho_{m} + \rho_{UR} + \sum \, \rho_{B i}, \, \, \, \, \, \, \,
p_{T} = p_{m} + p_{UR} + \sum \, p_{B i},  \label{total}
\end{equation}
where
\begin{eqnarray}
\rho_{m} &=& \rho_{m0} \left(\frac{a_{0}}{a}\right)^{3}
\,\,\,\,\,\,\,\,\,\,\,\, , \,\,\,\,\,\,\,\,\,\,\,\, p_{m} = 0  \nonumber \\
\rho_{UR} &=& \rho_{UR 0} \left(\frac{a_{0}}{a}\right)^{4}
\,\,\,\,\,\,\,\,\,\,\,\,\, , \,\,\,\,\,\,\,\,\,\,\,\,\, p_{UR} = \frac{1}{3}
\rho_{UR}  \nonumber \\
\rho_{B 1} &=& - \, 16 \alpha^{2}\, B_{0}^{4} \, \left(\frac{a_{0}}{a}%
\right)^{8} \,\, , \,\, p_{B 1} = \frac{5}{3} \rho_{B 1}  \nonumber \\
\rho_{B 2} &=& B_{0}^{2} \, \left(\frac{a_{0}}{a}\right)^{4}
\,\,\,\,\,\,\,\,\,\,\,\, , \,\,\,\,\,\,\,\,\,\,\,\, p_{B 2} = \frac{1}{3}
\rho_{B 2}  \nonumber \\
\rho_{B 3} &=& \, \frac{\mu^{2}}{4 B_{0}^{2}} \, \left(\frac{a}{a_{0}}%
\right)^{4} \,\,\,\,\,\,\,\, , \,\,\,\,\,\,\,\, p_{B 3} = - \, \frac{7}{3}
\, \rho_{B 3}  \nonumber \\
\rho_{B 4} &=& - \, \frac{\beta^{2}}{16 B_{0}^{4}} \, \left(\frac{a}{a_{0}}%
\right)^{8} \,\,\,\, , \,\,\,\, p_{B 4} =-\, \frac{11}{3} \, \rho_{B 4}.
\label{densidades}
\end{eqnarray}

The real evolution of the universe depends on the values of the six
constants presents in (\ref{densidades}). However, at least in a qualitative
manner, it is possible to distinguish five distinct cosmological stages just
looking the form of the terms in (\ref{densidades}). These five stages are:

\begin{itemize}
\item {The bouncing era: this era is governed by $\rho_{B 1}$ and it
dominates in the earliest universe (before nucleosynthesis). In
bouncing era the scale factor reaches a minimum value.}

\item {The radiation era: this era is governed by $\rho_{UR}$ with
$a(t)\sim t^{1/2}$. During this epoch occurs the primordial
nucleosynthesis.}

\item {The matter era: this era is dominated by $\rho_{m}$with
$a(t)\sim t^{2/3}$. It is expected that the structure formation
occurs in this epoch.}

\item {The acceleration era: in this period $\rho_{B 3}$ govern the cosmic
evolution. The term $\rho + 3 p$ becomes negative and the universe
begins to accelerate.}

\item {The re-bouncing era: this era is dominated by $\rho_{B 4}$.
The term $\rho + 3 p$ becomes positive once more and starts a new
decelerated phase. After a while, the $a(t)$ reach a maximum
and re-bounces entering in a collapsing phase.}
\end{itemize}

In the section \ref{sec-ComSce}, it will be shown explicitly that
the universe pass through of all the five stages.

\section{Setting the six constants}

In order to analyze the complete scenario it is necessary to
determine six constants where only five are independent\footnote{In
this work it is considered only Euclidean spatial section, i.e.
$\kappa=0$.}. Although five independent constants are a elevated
number of free parameters, this model describes all the evolution
for the universe, since the bounce until the re-bounce. It keeps
untouchable the main successes of the standard Big Bang model
(primordial nucleosynthesis, CMB generation, etc) and solve some
other problems like the initial singularity, the present
acceleration and the Big Rip.

The six parameters are divided in two distinct classes: the first one,
represented by $\alpha$, $B_{0}$ and $\rho_{UR0}$, acts on the primordial
universe; and the second one, represented by $\mu$, $\beta$ and $\rho_{m0}$,
affects only the recent universe. Because this difference, each class must
be fixed by independent manner. Let's start with the primordial constants.

\subsection{Constants for the primordial universe}

\label{Sec-CPU}

In this subsection it will be fixed or constrained the three primordial parameters namely
$\alpha$, $B_{0}$ and $\rho_{UR0}$. The most simple to determine is $%
\rho_{UR0}$, so let\'{}s start with it.

The constant $\rho_{UR0}$ represents the nowadays energy density for the all
ultra-relativistic particles. It is basically composed of radiation $%
\rho_{\gamma 0}$ and neutrinos $\rho_{\nu 0}$. For radiation, the precisely
CMB measurement \cite{PDG} shows that,

\begin{eqnarray}
\rho_{\gamma 0}=4.6 \times 10^{-5} \rho_{c} \text{\, \,with\, \,}
\rho_{c}\equiv\frac{8\pi G}{3 H_{0}^{2}}=1.9 \times 10^{-9}h^{-2} \text{erg}
\nonumber
\end{eqnarray}
where $h\simeq0.72$.

Theoretical considerations for a massless neutrinos \cite{WeinCosm} imply in

\begin{eqnarray}
\rho_{\nu 0}=6 \times \frac{7}{16} \times \left(\frac{4}{11}\right)^{4/3}
\rho_{\gamma 0}=3.1 \times 10^{-5} \rho_{c}.  \nonumber
\end{eqnarray}

Therefore, the nowadays ultra-relativistic energy density is

\begin{equation}
\rho_{UR0}=7.7 \times 10^{-5} \rho_{c} .  \label{Rho_UR0}
\end{equation}

The next parameter it should be fix is $B_{0}$. As could be seen in (\ref%
{densidades}), the constant $B_{0}$ acts directly only in the
primordial universe through $\rho_{B 2}$. Nevertheless, its
numerical value is necessary to determine the $\alpha$, $\beta$ and $\mu$ constants. Thus, $%
B_{0}$ affects not only the primordial universe but also, in the indirectly
way, the present universe.

Physically the parameter $B_{0}$ is related with a cosmological magnetic
field. Supposing the existence of a large scale magnetic field, its
background mean value $\overline{B}_{0}$ must be $\overline{B}_{0}<10^{-9}$ $%
G$ \cite{Mag}. In the $cgs$ system, the energy density related with this
magnetic field is

\begin{equation}
\rho _{\bar{B}_{0}}=\frac{\bar{B}_{0}^{2}}{8\pi }=4\times 10^{-20}b^{2}\text{
}erg\text{,}  \label{Ene_Mag}
\end{equation}
where $b$ is just a parametrization ($b\leq1$).

On the other hand, the present energy density for the cosmological magnetic
field $\rho _{B2_{0}}$ is given by

\[
\rho _{B2_{0}}=\frac{\bar{F}}{4}=B_{0}^{2}.
\]

So, the comparison between the last equation and (\ref{Ene_Mag}) results in

\begin{equation}
B_{0}^{2}=4\times 10^{-20}b^{2}\text{ }erg=1.1\times 10^{-11}b^{2}\text{ }%
\rho_{c}.  \label{B0^2}
\end{equation}

Replacing (\ref{B0^2}) and (\ref{Rho_UR0}) in (\ref{densidades}) a first
relevant result arises: $\rho_{B2}$ is always subdominant, i.e. the Maxwell
term concerning to cosmological magnetic field is not important to the
universe evolution. In fact, even the actual cosmological experiments
including CMB anisotropy observation performed by WMap satellite do not have
accuracy to measure a cosmological magnetic field. Perhaps, the future
experiments of CMB anisotropy (e.g. Planck satellite) are able to detected
it.

The last parameter necessary to fix is $\alpha$. This parameter
rules the bouncing stage, hence it is concerning at the very early
universe. These features make difficult to precisely determine
$\alpha$. Therefore, instead of to fix it we just establish a range
of validity for $\alpha$. The maximum and minimum allowed values are
obtained through of two physical conditions: the bounce must occurs
considerably before primordial nucleosynthesis but it should not
extrapolate the validity of classical gravitation. The first
condition implies in $\left\vert \rho _{UR}\left( a_{nuc}\right)
\right\vert \ll \rho _{UR}\left( a_{nuc}\right)$ i.e.,

\[
\alpha ^{2}\ll \frac{\rho _{\gamma 0}}{64B_{0}^{4}}\left( \frac{a_{nuc}}{%
a_{0}}\right) ^{4}.
\]

Replacing (\ref{Rho_UR0}) and (\ref{B0^2}) in the last equation it follows
\[
\alpha ^{2}\ll \frac{5.\,\allowbreak 94\times 10^{15}}{b^{2}}\left( \frac{%
T_{\gamma 0}}{T_{\gamma nuc}}\right) ^{4}\rho _{c}^{-1},
\]
where the scale factor was converted to temperature through of the relation $%
a\sim T^{-1}_{\gamma}$.

Using $T_{\gamma 0}\simeq 2\times 10^{-10} MeV$ and supposing that the
nucleosynthesis took place around $1$ $MeV$ we finally obtain
\begin{equation}
\alpha ^{2}\ll \frac{9.\,\allowbreak 5\times 10^{-24}}{b^{2}}\rho _{c}^{-1}.
\label{Max_Lim}
\end{equation}

For consistence with second condition the bounce should happen below
the Planck scale, i.e. $T_{\gamma Bou}\ll 10^{-19}GeV$. In this
model, the bouncing occurs when $\rho_{UR}+\rho_{B1}\simeq0$ which
implies in
\[
\left( \frac{a_{Bou}}{a_{0}}\right) ^{4}=\left( \frac{T_{\gamma 0}}{%
T_{\gamma Bou}}\right) ^{4}=\frac{64B_{0}^{4}}{\rho _{\gamma 0}}\alpha ^{2}.
\]
So,
\[
\left( \frac{T_{\gamma 0}}{T_{\gamma Bou}}\right) ^{4}\gg 1.\,\allowbreak
6\times 10^{-127},
\]
and hence
\begin{equation}
\alpha ^{2}\gg \frac{9.\,\allowbreak 5\times 10^{-112}}{b^{2}}\rho _{c}^{-1}.
\label{Min_Lim}
\end{equation}

At last, using (\ref{Max_Lim}) and (\ref{Min_Lim}), it is obtained the range
of validity for the parameter $\alpha$:

\begin{equation}
\frac{9.\,\allowbreak 5\times 10^{-112}}{b^{2}}\rho _{c}^{-1}\ll \alpha
^{2}\ll \frac{9.\,\allowbreak 5\times 10^{-24}}{b^{2}}\rho _{c}^{-1}.
\label{Num_alfa}
\end{equation}

The term $\alpha ^{2}F^{2}$ at the Lagrangian could be interpreted as an
one-loop quantum correction in the infrared regime. Indeed, it correction
was obtained first by Euler-Heisenberg \cite{Euler}. In this
context, $\alpha ^{2}$ values

\begin{eqnarray}
\alpha ^{2}=\left( \frac{1}{90}\right) \left( \frac{1}{137}\right)
^{2}\left( \frac{\hbar }{m_{e}c}\right) ^{3}\left( \frac{1}{m_{e}c^{2}}%
\right) \Rightarrow  \nonumber \\
\alpha ^{2}\simeq 4.\,\allowbreak 2\times 10^{-32}\frac{cm^{3}}{erg}
=3.\,\allowbreak 68\times 10^{-40}\rho _{c}^{-1}.  \label{EH}
\end{eqnarray}

Comparing (\ref{EH}) with (\ref{Num_alfa}) it could be seen that the both
results are consistent. So, in this scenario, the bouncing stage was
generated by quantum electrodynamics effects.

\subsection{Constants for the recent universe}

In this subsection, we intend to determine the three remain constants namely $%
\mu$, $\beta$ and $\rho_{m0}$. Differently for the three previous
parameters, these constants are relevant at the same time (present
universe). For this reason, they should be set simultaneously.

The procedure that was adopted is based on the fit of the model using
supernova data. This procedure concerns only to recent universe, so it could
be neglected the terms $\rho_{B1}$, $\rho_{UR}$ and $\rho_{B2}$. In this
case, the luminosity distance $d_{L}$ is written as
\begin{equation}
d_{L}\left( z\right) \equiv \frac{d_{h}}{H_{0}}=\left( 1+z\right)
\int\limits_{0}^{z}\frac{d\bar{z}}{H\left( \bar{z}\right) }  \label{dL}
\end{equation}
where the Hubble function $H(z)$ is given by
\begin{equation}
\frac{H(z)}{H_{0}}= \sqrt{\Omega _{m0}\left( 1+z\right) ^{3}+\Omega _{\mu
0}\left( 1+z\right) ^{-4}+\Omega _{\beta 0} \left( 1+z\right) ^{-8}}
\label{H_z}
\end{equation}
with the three $\Omega$'s defined as
\begin{equation}
\Omega _{m0}\equiv \frac{\rho _{m0}}{\rho _{c}}\text{,\ }\Omega _{\mu
0}\equiv \frac{\mu ^{2}}{4B_{0}^{2}\rho _{c}}\text{\ and\ }\Omega _{\beta
0}\equiv -\frac{\beta ^{2}}{16B_{0}^{4}\rho _{c}}.  \label{Omega_Definitions}
\end{equation}
For $z=0$, (\ref{H_z}) implies in $\Omega _{m0}+ \Omega _{\mu 0}+\Omega
_{\beta 0}=1$. Thus, for the present universe, the model has only two free
parameters.

The set of SNIa which was used in the fit is the Union set \cite{Union}.
This set collects the mainly sets of supernovas which was obtained during
the last twelve years. It is constituted of 414 SN divided in 13 sub-sets
where all of them were analyzed through of the same procedure. Based on this
procedure, the authors excluded 107 SN, so only 307 supernovas were regarded
true standard candles. It is with these 307 SN which is fitted the two
parameters.

The supernova data are usually assumed as Gaussian, hence it could be used
the statistical likelihood method to determine the free parameters. The
standard procedure consist in to build and minimize a $\chi ^{2}$ through
the fit of the model's parameter. Following the steps point out in \cite%
{Union}, let's start the construction of $\chi ^{2}$ regarding only the
statistical errors:

\begin{equation}
\chi ^{2}=\sum\limits_{i}\frac{\left[ \mu _{r}\left(
m_{i},s_{i},c_{i}|M,\alpha ,\beta \right) -\mu \left( z_{i}|\Omega
_{m0},\Omega _{\mu 0}\right) -\mathcal{M}\right] ^{2}}{\sum%
\limits_{jk}C_{jk}e_{j}e_{k}+\sigma _{tot}^{2}+\sigma _{int}^{2}},
\label{Chi^2 geral}
\end{equation}
with
\[
\mu \left( z_{i}|\Omega _{m0},\Omega _{\mu 0}\right) =5\log d_{h}\left(
z_{i}|\Omega _{m0},\Omega _{\mu 0}\right)
\]
and
\begin{equation}
\mu _{r}=m_{i}-M+\alpha \left( s_{i}-1\right) -\beta c_{i},  \label{muri}
\end{equation}
where $m_{i}$ is the $B$-band peak magnitude, $M$ is the absolute magnitude,
$s_{i}$ represents the stretch in supernova lightcurve, $c_{i}$ represents
the color variation between $B$ and $V$ bands, $\alpha$ is the stretch
parameter, $\beta$ is the color parameter and $\mathcal{M}$ is a additive
factor which include constants like $H_{0}$. The $C_{jk}$ matrix is the
statistical covariant matrix, $\sigma _{tot}$ is related with astrophysical
dispersion and $\sigma _{int}$ represents the intrinsical dispersion.

The term $\sum\nolimits_{jk}C_{jk}e_{j}e_{k}$ represents the uncertainties
associated with $m_{i},s_{i}$ and $c_{i}$ where $e_{j}=\left( 1,\alpha
,-\beta \right)$. In principle $\alpha $ and $\beta $ are free parameters
which must be fitted for each specific cosmological model. Nevertheless, the
authors in \cite{Union} found out that $\alpha $ and $\beta $ are almost
independent of cosmological parameters. So, it is assumed that $\alpha $ and
$\beta$ are constants\footnote{%
This approximation engender an new source of error which will be regarded as a
systematic error.}, the uncertainties of $m_{i},s_{i}$ and $c_{i}$ could be
directly passed to $\mu_{r_{i}}$. In this case, the $\chi ^{2}$ is written
as
\begin{equation}
\chi ^{2}\left( \Omega _{m0},\Omega _{\mu 0},\mathcal{M}\right)
=\sum\limits_{i=1}^{N}\frac{\left( \mu _{r_{i}}-\mu \left( z_{i}|\Omega
_{m0},\Omega _{\mu 0}\right) -\mathcal{M}\right) ^{2}}{\sigma _{\mu
_{r_{i}}}^{2}+\sigma _{tot}^{2}+\sigma _{int}^{2}},  \label{Chi^2 esta}
\end{equation}%
where the parameter $M$ was embedded in $\mathcal{M}$. In \cite{Union} is
available the list with $z$, $\mu _{r_{i}}$ and $\sigma_{\mu _{r_{i}}}$ for
each one of the 307 SNIa.

Beside $\sigma_{\mu _{r_{i}}}$, there are other three gaussian sources of
uncertainties: the first one is the dispersion due gravitational lensing,
the second one is the uncertainty in the Milk-Way dust extinction correction
and the last one is the uncertainty due peculiar velocity in the host galaxy.

\begin{itemize}
\item Gravitational lensing decreases the mode of the brightness
distribution and causes increased dispersion in the Hubble diagram at high
red-shift \cite{Union,Sas,Hol_Lin}. The both strong and weak leasing effects
engender a dispersion of $0.093\times z$ in magnitude \cite{Hol_Lin}, and if
the statistical of SNIa is large enough this uncertainty could be considered
as a Gaussian error.

\item All the SNIa light-curve data were corrected for extinction caused by
our Galaxy. Nevertheless, these corrections contain statistical and
systematic errors which act directly on the magnitude value. These
statistical uncertainties will be neglected because three reasons: they only
concern the SN with $z<0.2$, and these ones represent approximately $20 \%$
of total set; out of Galactic plane these uncertainties value as a rule $%
\sim 0.01$; the implementation of these uncertainties is very complicated
because the extinction and its errors change with the Galactic coordinates.
The systematic uncertainties are treated later on.

\item As point out in several papers \cite{Union,Astier,Riess1998,Perm}, the
peculiar velocity in the host galaxy causes a error $\sigma _{v}$ which
varies between $100$ and $500$ $Km/s$. In this paper, we adopt $\sigma
_{v}=300$ $km/s$. So, the uncertainty propagated to magnitude due the
peculiar velocities is
\[
\sigma _{\mu _{v}}=\frac{5}{d_{h}\ln 10}\sigma _{d_{h}}\simeq \frac{2.172}{z}%
\frac{\sigma _{v}}{c},
\]
where $\sigma _{d_{h}}=\frac{\sigma _{v}}{c}=0.001$.
\end{itemize}

These three sources of uncertainty are encapsulated in $\sigma _{tot}$. Thus,
the discussion above results in
\begin{equation}
\sigma _{tot}^{2}=\sigma _{les}^{2}+\sigma _{\mu _{v}}^{2}=\left(
0.093z\right) ^{2}+\left( \frac{2.172}{z}\frac{\sigma _{v}}{c}\right) ^{2}.
\label{sig_tot}
\end{equation}

The last source of statistical error which was considered is the uncertainty
produced by an intrinsical dispersion $\sigma _{int}$. This intrinsical
dispersion is related about non corrected and/or unknown errors.

To estimate $\sigma _{int}$ it is performed a first fit with a small initial
value ($\sim 0.01 mag$). Then it is determined a new $\sigma _{int}$
imposing $\chi^{2}=1$. And the last step consists in perform the same
routine once more to obtain a more accurate and definitive $\sigma _{int}$.
During this procedure it is not used $\sigma _{tot}$ because the intrinsical
dispersion is related only with the "main" data namely $\left\{ z,\mu
,\sigma _{\mu }\right\}$.

\subsubsection{Systematic uncertainties}

The systematic uncertainties are introduced at the same way which was
performed in \cite{Union}. The method consists in to introduce a gaussian
distribution with a $\Delta M_{s}$ parameter for each present systematic
uncertainty $\sigma _{\Delta M_{s}}$. The new parameters get in additively
in the modulus of distance, i. e.
\[
\mu \left( z|\Omega _{m0},\Omega _{\mu 0}\right) \rightarrow \mu \left(
z|\Omega _{m0},\Omega _{\mu 0}\right) +\sum\limits_{s}\Delta M_{s},
\]
and the previous likelihood changes to
\[
\mathcal{L} _{prev} \left(\Omega _{m0},\Omega _{\mu 0}\right) \rightarrow
\mathcal{L} _{post} \left(\Omega _{m0},\Omega _{\mu 0},\Delta M_{s}\right)
\]
where
\[
\mathcal{L} _{post} \left(\Omega _{m0},\Omega _{\mu 0},\Delta
M_{s}\right)\equiv \mathcal{L} _{prev} \left(\Omega _{m0},\Omega _{\mu
0}\right) \prod\limits_{s} e^{{-\frac{\Delta M^{2}_{s}}{2\sigma _{\Delta
M_{s}}^{2}}}}.
\]

In this kind of construction, we are supposing that a specific source of
systematic error leads to a "variation" in the modulus of distance. And the
importance of its variation is determined through of a gaussian distribution
where the standard deviation is identified with the systematic uncertainty.

At the Union set, there are two kinds of systematic errors: one which
depends on the specific sub-set (e. g. due to observational effects) and the
other which concerns to all set of supernovas (e. g. due to astrophysical or
fundamental calibration effects). The systematic uncertainties common to all
samples can be absorbed in the definition of absolute magnitude. On the
other hand, it is expected a difference between the nearby ($z\sim 0.05$)
and distant ($z\sim 0.5$) supernovas. So, we can cast the common systematic
uncertainties to a uncertainty in the difference $\Delta M$ between absolute
magnitudes of close and distant SNs. Following the steps discussed in \cite%
{Union}, it is defined $z_{div}=0.2$ as the point which splits the SNs in
nearby and farther objects. Beside $\Delta M$, we introduce a set of 13
parameters $\Delta M_{s}$ which represents the systematic uncertainties for
each one of the 13 sub-sets.

The complete likelihood for the model is given by
\begin{equation}
\mathcal{L} _{post}=e^{-\frac{\chi _{post}^{2}}{2}},  \label{like_compl}
\end{equation}
where
\begin{eqnarray}
\chi _{post}^{2} &=& \left[ \sum\limits_{s=1}^{13} \left[ \sum%
\limits_{i=1}^{N_{s}}\frac{\left( \mu _{r_{is}}-\mu +\Delta M^{\ast }+\Delta
M_{s}-\mathcal{M}\right) ^{2}}{\sigma _{\mu _{r_{is}}}^{2}+\sigma
_{tot}^{2}+\sigma _{int}^{2}} \right] \right. +  \nonumber \\
&& + \left. \frac{\Delta M_{s}^{2}}{\sigma _{\Delta M_{s}}^{2}}\right] +
\frac{\Delta M^{2}}{\sigma _{\Delta M}^{2}}.  \label{chi_completo}
\end{eqnarray}
with
\[
\Delta M^{\ast }=\left\{
\begin{array}{c}
\Delta M\text{ \ \ \ for \ \ }z>0.2 \\
0\text{ \ \ \ \ \ \ \ for \ \ }z\leq 0.2%
\end{array}
\right. .
\]
The Union set was split in the sum of the 13 sub-set where each sub-set has $%
N_{s}$ samples. Thus, comparing (\ref{Chi^2 esta}) with (\ref{chi_completo})
it follows that $\sum\limits_{s=1}^{13}N_{s}=N$.

The systematic errors arise from seven distinct sources:

\begin{itemize}
\item Errors due the fixation of $\alpha $ and $\beta $ parameters.

\item Errors due the possible samples contamination.

\item Errors due the models of supernovas light curve and $K$-corrections.

\item Error due photometric zero point calibration.

\item Error due a possible variation of Malmquist bias with the red-shift.

\item Systematic errors due gravitational lensing effects.

\item Systematic error due the normalization in the corrections for Galactic
extinction maps.
\end{itemize}

Taking account all these effects, the authors of \cite{Union} reached the
following uncertainties:
\[
\sigma _{\Delta M}=0.04\text{ \ \ and \ \ }\sigma _{\Delta M_{s}}=0.033\text{
\ for all }s.
\]%
So, the error related with each one of the 13 sub-sets are the same.

\subsubsection{Results}

First, we take account only the statistical uncertainties. Using the data
presents in \cite{Union}, it is performed a numerical calculation to
determine the cosmological parameters. The procedure could be summarized in
five steps:

\begin{enumerate}
\item Construction of $\mathcal{L} \left( \Omega _{m0},\Omega _{\mu 0},%
\mathcal{M} \right)$ not taking account $\sigma _{tot}$.

\item Marginalization of $\mathcal{M}$ parameter\footnote{%
We are interested only in cosmological parameters.}.

\item Calculation of $\sigma _{int}$.

\item Construction of the new likelihood taking account $\sigma _{tot}$ and
the $\sigma _{int}$.

\item And finally, the minimization of $\chi ^{2}\left( \Omega _{m0},\Omega
_{\mu 0} \right)$.
\end{enumerate}

\begin{figure}[ht]
\begin{center}
\includegraphics[height=7.0cm, width=6.5cm]{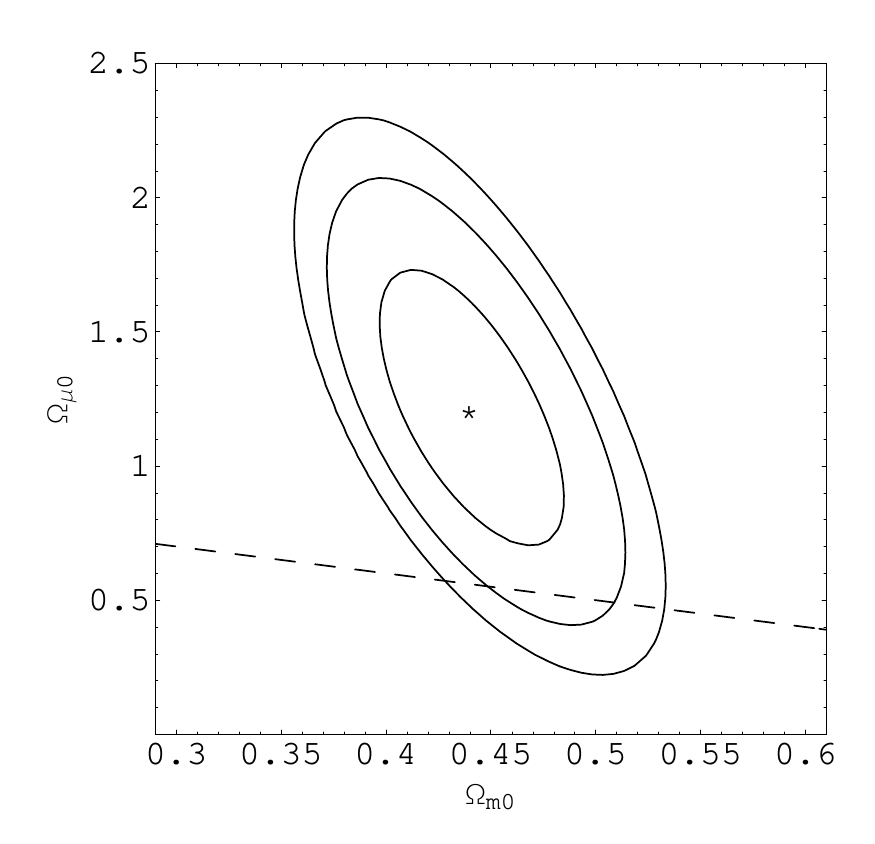}
\end{center}
\caption{Parametric contour plot for $\Omega_{m0}\times \Omega_{\protect\mu %
0}$ taking account only the statistical uncertainties. From inside to
outside the contour lines correspond respectively to 68\%, 95\% and 99\%
confidence level. The dashed line represents $\Omega_{\protect\beta 0}=0$.}
\label{figura1}
\end{figure}

The results are shown in figure \ref{figura1}.


The best fit for $\Omega_{m0}$ or $\Omega_{\mu 0}$ could be established
marginalizing in all the other parameters. Performing these marginalization
we get in the following results:
\begin{eqnarray}
\Omega _{m0} =0.44_{-0.03}^{+0.03} \text{ \ \ and \ \ } \Omega _{\mu 0}
=1.20_{-0.33}^{+0.35} \text{ }.  \label{OmegaMmu1}
\end{eqnarray}

Finally, it is determined $\Omega_{\beta 0}$ through of the equation $\Omega
_{m0}+ \Omega _{\mu 0}+\Omega _{\beta 0}=1$. Thus,
\begin{eqnarray}
\Omega _{\beta 0} =-0.64_{+0.33}^{-0.35} \text{ }.  \label{Omegabeta1}
\end{eqnarray}

The next step is to include the systematic uncertainties in the procedure.
Analogously to the first case, it is performed an analytic marginalization at the
\textquotedblleft non-cosmological parameters\textquotedblright \, namely $%
\mathcal{M}$, $\Delta M$ and the set of $\Delta M_{s}$. Again, the method is
straightforward but the calculation is much more cumbersome than the
previous case.

Using the same steps described earlier, it is obtained the following result:


\begin{figure}[ht]
\begin{center}
\includegraphics[height=7.0cm, width=6.5cm]{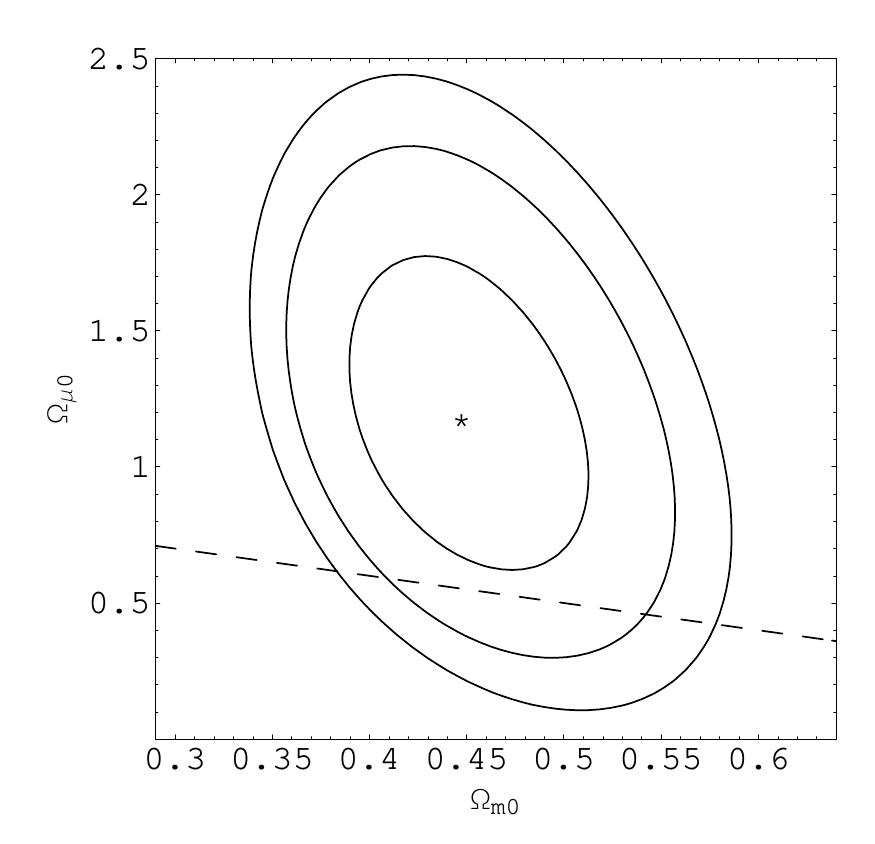}
\end{center}
\caption{Parametric contour plot for $\Omega_{m0}\times \Omega_{\protect\mu %
0}$ taking account the statistical and systematic uncertainties. From inside
to outside the contour lines correspond respectively to 68\%, 95\% and 99\%
confidence level. The dashed line represents $\Omega_{\protect\beta 0}=0$.}
\label{figura2}
\end{figure}

For this case, the best fits for $\Omega_{m0}$ and $\Omega_{\mu 0}$ are
\begin{eqnarray}
\Omega _{m0} =0.448_{-0.039}^{+0.042} \text{ \ \ and \ \ } \Omega _{\mu 0}
=1.17_{-0.37}^{+0.39} \text{ },  \label{OmegaMmu2}
\end{eqnarray}
and the calculated value for $\Omega_{\beta 0}$ is
\begin{eqnarray}
\Omega _{\beta 0} =-0.62_{+0.37}^{-0.39} \text{ }.  \label{Omegabeta2}
\end{eqnarray}

As expected, the inclusion of systematic uncertainties increases the
uncertainties at the cosmological parameters.

\subsubsection{Including a prior in $\Omega _{m0}$}

Measurements of X-ray gas mass fraction $f_{gas}$ done in galaxy clusters
allow to establish excellent values to the ratio $\Omega _{b0}/\Omega _{m0}$
\cite{Schindler,Allen}. On the other hand, primordial nucleosynthesis
determine with good accuracy the actual barion density $\Omega _{b0}$. So,
join these two facts it is possible to obtain an excellent estimation for $%
\Omega _{m0}$. Based on this estimation, we intend to build a gaussian prior
to $\Omega _{m0}$ as follows:
\[
P=Exp\left[ -\frac{\left( \Omega _{m0}-\bar{\Omega}_{m0}\right) ^{2}}{%
2\sigma _{\bar{\Omega}_{m0}}^{2}}\right],
\]
where $\bar{\Omega}_{m0}$ and its uncertainty are obtained through of gas
mass fraction procedure.

According to \cite{Allen}, the $f_{gas}$ is linked with $\Omega _{b0}/\bar{%
\Omega}_{m0}$ as shown below:
\begin{equation}
f_{gas}^{\Lambda CDM}\left( z\right) =\frac{KA\gamma b\left( z\right) }{%
1+s\left( z\right) }\left( \frac{\Omega _{b0}}{\bar{\Omega}_{m0}}\right) %
\left[ \frac{d_{A}^{\Lambda CDM}\left( z\right) }{d_{A}\left( z\right) }%
\right] ^{\frac{3}{2}},  \label{fgas}
\end{equation}
where $d_{A}$ is the angular diameter distance given by
\begin{equation}
d_{A}\left( z\right) =\frac{1}{\left( 1+z\right) }\int\limits_{0}^{z}\frac{dz%
}{H\left( z\right) }.  \label{dA}
\end{equation}
The $f_{gas}^{\Lambda CDM}$ is the gas mass fraction for $\Lambda CDM$ model
(the reference model) with the following parameters: $\Omega _{m0}=0.3$, $%
\Omega _{\Lambda} =0.7$ and $h=0.7$. The other factors which appear in (\ref%
{fgas}) are described below:

\begin{itemize}
\item $K$ is a constant that parameterizes uncertainties in the accuracy of
the instrument calibration and X-ray modeling. Its value is $K=1\pm 0.1$.

\item $\gamma $ is related with the non-thermic pressure support in the
clusters. Its value is $1.0<\gamma <1.1$.

\item $b\left( z\right)$ is the depletion factor, i.e. the ratio by which
the baryon fraction is depleted with respect to the universal mean. It is
modeled by $b\left( z\right) =b_{0}\left( 1+\alpha _{b}z\right) $ where $%
b_{0}=0.83\pm 0.04$ and $-0.1<\alpha _{b}<0.1$.

\item The function $s\left( z\right) =s_{0}\left( 1+\alpha _{s}z\right)$
describes the baryonic mass fraction in stars where $s_{0}=\left( 0.16\pm
0.05\right) h_{70}^{1/2}$ and $-0.2<\alpha _{s}<0.2$.

\item The $A$ factor take account the change in the subtended angle of the
clusters due the difference between $\Lambda CDM$ model and the others
models. According to authors in \cite{Allen}, this effect could be neglect
for all red-shifts of interest, i.e. $A=1$.
\end{itemize}

To estimate $\bar{\Omega}_{m0}$, it is used a set of six clusters with low
red-shifts (see table \ref{table})

\begin{table}[ht]
\begin{tabular}{|c|c|c|}
\hline
Cluster classification & $z$ & $f_{gas}^{\Lambda CDM}h_{70}^{1.5}$ \\ \hline
Abell 1795 & $0.063$ & $0.1074\left( 0.0075\right) $ \\ \hline
Abell 2029 & $0.078$ & $0.1117\left( 0.0042\right) $ \\ \hline
Abell 478 & $0.088$ & $0.1211\left( 0.0053\right) $ \\ \hline
PKS0745-191 & $0.103$ & $0.1079\left( 0.0124\right) $ \\ \hline
Abell 1413 & $0.143$ & $0.1082\left( 0.0058\right) $ \\ \hline
Abell 2204 & $0.152$ & $0.1213\left( 0.0116\right) $ \\ \hline
\end{tabular}%
\caption{The red-shift and $f_{gas}$ for the six galaxy clusters. The data
were taken from table 3 of reference \protect\cite{Allen}.}
\label{table}
\end{table}

As we are considering only clusters at low red-shift, it can be neglected the
dependence on $z$ of the functions $b\left( z\right)$ and $s\left( z\right)$%
. Thus, inverting (\ref{fgas}) it follows:

\begin{equation}
\bar{\Omega}_{m0}\simeq \frac{Kb_{0}}{\left( 1+s_{0}\right) f_{gas}^{\Lambda
CDM}} \left[ \frac{d_{A}^{\Lambda CDM}\left( z\right) }{d_{A}\left( z\right)
} \right] ^{\frac{3}{2}}\Omega _{b0}.  \label{OM_prior}
\end{equation}
Face on the other uncertainties, the variation of gamma could be neglected,
i.e. $\gamma =1$. Using primordial nucleosynthesis models and data of
deuterium abundance, it is possible to derive the value for $\Omega _{b0}$
with good accuracy \cite{Kirkman}:
\begin{equation}
\Omega _{b0}h^{2}=0.0214\pm 0.002\text{ \ \ \ with \ \ \ }h=0.7\pm 0.1,
\end{equation}
where the value of $h$ was determined in \cite{Fredmann}.

The procedure to determine $\bar{\Omega}_{m0}$ could be summarized in the
following steps:

\begin{enumerate}
\item Fit of parameters using the set of SNIa without priors. The steps were
discussed previously.

\item Calculation of $d_{A}(z)$ using the previous fit.

\item Determination of $\bar{\Omega}_{m0}$ and its corresponding uncertainty
for each one of the six clusters.

\item Calculation of weighted mean and the mean uncertainty for $\bar{\Omega}%
_{m0}$.

\item Finally, it repeats the previous step but now using the prior for $%
\Omega_{m0}$.
\end{enumerate}

These steps was performed some times until $\bar{\Omega}_{m0}$ stabilize in
\begin{equation}
\bar{\Omega}_{m0} =0.278\pm 0.038.  \label{Omega_bar}
\end{equation}

\begin{figure}[ht]
\begin{center}
\includegraphics[height=7.0cm, width=6.5cm]{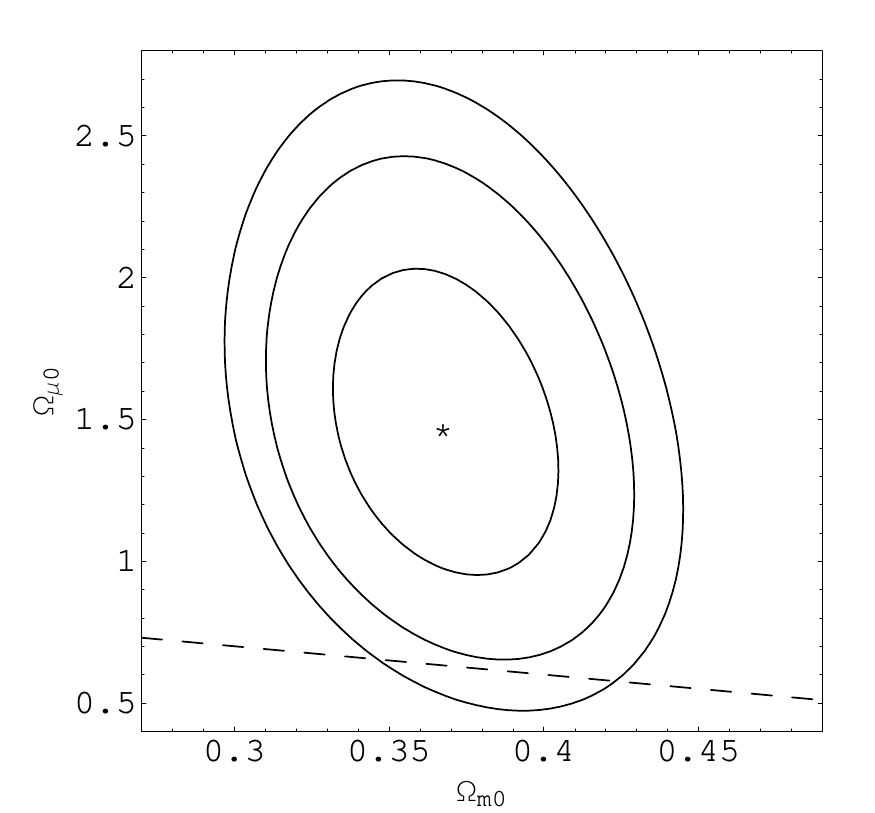}
\end{center}
\caption{Parametric contour plot for $\Omega_{m0}\times \Omega_{\protect\mu %
0}$ taking account the statistical and systematic uncertainties and a prior
in $\Omega_{m0}$. From inside to outside the contour lines correspond
respectively to 68\%, 95\% and 99\% confidence level. The dashed line
represents $\Omega_{\protect\beta 0}=0$.}
\label{figura3}
\end{figure}

The complete result including the statistical and systematic uncertainties
and the prior in $\Omega_{m0}$ is shown in figure \ref{figura3}. The best
fits for $\Omega_{m0}$ and $\Omega_{\mu 0}$ are
\begin{eqnarray}
\Omega _{m0} =0.367_{-0.023}^{+0.026} \text{ \ \ and \ \ } \Omega _{\mu 0}
=1.47_{-0.34}^{+0.37} \text{ },  \label{OmegaMmu3}
\end{eqnarray}
and the calculated value for $\Omega_{\beta 0}$ is
\begin{eqnarray}
\Omega _{\beta 0} =-0.84_{+0.34}^{-0.37} \text{ }.  \label{Omegabeta3}
\end{eqnarray}


It is interesting determine explicitly the values of $\mu$ and $\beta$.
Thus, replacing (\ref{OmegaMmu3}), (\ref{Omegabeta3}) and (\ref{B0^2}) in (%
\ref{Omega_Definitions}) follows:

\begin{eqnarray*}
\mu &=&\sqrt{4B_{0}^{2}\rho _{c}\Omega _{\mu 0}}=8.0\times 10^{-6}b\text{ }%
\rho _{c}. \\
\beta &=&\sqrt{-16B_{0}^{4}\rho _{c}\Omega _{\beta 0}}=3.\,\allowbreak
98\times 10^{-11}b^{2}\text{ }\rho _{c}^{3/2}.
\end{eqnarray*}

Since $\rho _{c}\sim 10^{-9} erg$, the values of $\mu$ and $\beta$
are extremely small and undetectable in the classical earth
experiments. Other interesting feature of this model is that the
value of $\Omega _{m0}$ is greater than the majority models presents
in literature. Even with the inclusion of the prior in $\Omega
_{m0}$ the difference still remains. It is possible that this
feature are related with the eminent deceleration provide by $\beta
^{2}/F^{2}$ term. A last point important to comment is that with the
inclusion of this prior, the region where $\Omega_{\beta 0}>0$
(below to the dashed line) is excluded in $2 \sigma$. It is a
important feature because with positive $\Omega_{\beta 0}$ the model
does not engender the re-bounce.

\section{The complete scenario \label{sec-ComSce}}

Before to discuss the complete scenario, it is necessary, based on
the previous section, choose the values for the six constant of this
model. For
the three constants associated with the primordial universe were selected (\ref%
{Rho_UR0}) for $\rho_{UR0}$, (\ref{B0^2}) with $b=1$ for $B_{0}^{2}$ and $%
\alpha ^{2}= 10^{-64}\rho _{c}^{-1}$ suitable with (\ref{Num_alfa}). For the
three constants related with the recent universe were chosen the best fit for $%
\Omega_{m0}$, $\Omega_{\mu 0}$ and $\Omega_{\beta 0}$ given by (\ref%
{OmegaMmu3}) and (\ref{Omegabeta3}). This best fit is the most convenient
choice to characterize the complete scenario because it incorporates more
observational cosmological features than the others two. Nevertheless, using
another best fit the qualitative results discussed below remain the
same.

\subsection{The bouncing period and the primordial acceleration}

\label{Sec-Boun}

Near to bounce, the Friedmann equations could be approximated by,

\[
\left( \frac{\dot{a}}{a}\right) ^{2}\simeq H_{0}^{2}\left[ \Omega
_{UR0}\left( \frac{a_{0}}{a}\right) ^{4}-\Omega _{\alpha 0}\left( \frac{%
a_{0} }{a}\right) ^{8}\right]
\]
and
\[
\frac{\ddot{a}}{a}\simeq -H_{0}^{2}\left[ \Omega _{UR0}\left( \frac{a_{0}}{a}
\right) ^{4}-3\Omega _{\alpha 0}\left( \frac{a_{0}}{a}\right) ^{8}\right]
\]
where
\[
\Omega _{UR0}=\frac{\rho _{UR0}}{\rho _{c}}\text{ \ and \ }\Omega _{\alpha
0}=\frac{16\alpha ^{2}B_{0}^{4}}{\rho _{c}}.
\]

The conditions for bouncing
\[
\dot{a}_{Bou}=0 \text{ \, \, and \, \, } \ddot{a}_{Bou}>0, \nonumber
\]
implies in
\[
\left( \frac{a_{Bou}}{a_{0}}\right) ^{4}=\frac{\Omega _{\alpha 0}}{\Omega
_{UR0}},
\]
which allows rewrite de Friedmann equations as
\begin{equation}
\left( \frac{\dot{a}}{a}\right) ^{2}\simeq H_{0}^{2}\frac{\Omega _{UR0}^{2}}{%
\Omega _{\alpha 0}}\left[ \left( \frac{a_{Bou}}{a}\right) ^{4}-\left( \frac{%
a_{Bou}}{a}\right) ^{8}\right]  \label{Fried_Bou3}
\end{equation}
and
\begin{equation}
\frac{\ddot{a}}{a}\simeq -H_{0}^{2}\frac{\Omega _{UR0}^{2}}{\Omega _{\alpha
0}}\left[ \left( \frac{a_{Bou}}{a}\right) ^{4}-3\left( \frac{a_{Bou}}{a}%
\right) ^{8}\right].  \label{Fried_Bou4}
\end{equation}
The two previous equations clearly show the existence of bounce.
Using the numerical values for $\alpha ^{2}$, $B_{0}^{2}$ and
$\rho_{UR0}$, it is possible determine the minimum for the scale
factor:
\begin{equation}
a_{Bou}\simeq \left( \frac{\Omega _{\alpha 0}}{\Omega _{UR0}}\right)
^{1/4}a_{0}\simeq 7.08\times 10^{-21}a_{0},  \label{Numboun}
\end{equation}
where in this calculation the contribution of non-relativistic matter
was neglected.

The solution for (\ref{Fried_Bou3}) was studied in \cite{Aline} and has the
following form:
\begin{equation}
a(t)=a_{Bou}\sqrt[4]{\left( 2Qt\right) ^{2}+1} .  \label{Sol_Bou}
\end{equation}
Here the initial condition $a(0)=a_{Bou}$ was used. The time scale $Q$ is
defined as%
\[
Q\equiv \frac{H_{0}\Omega _{UR0}}{\sqrt{\Omega _{\alpha 0}}}=4.1\times
10^{20}s^{-1},
\]
with $H_{0}=72$ $Km/s.Mpc$ - see \cite{Fredmann}. This solution is
valid throughout the range in which the non-relativistic matter is
negligible, e. g. from bounce to primeval nucleosynthesis.

\begin{figure}[ht]
\begin{center}
\includegraphics[height=6.0cm, width=7.2cm]{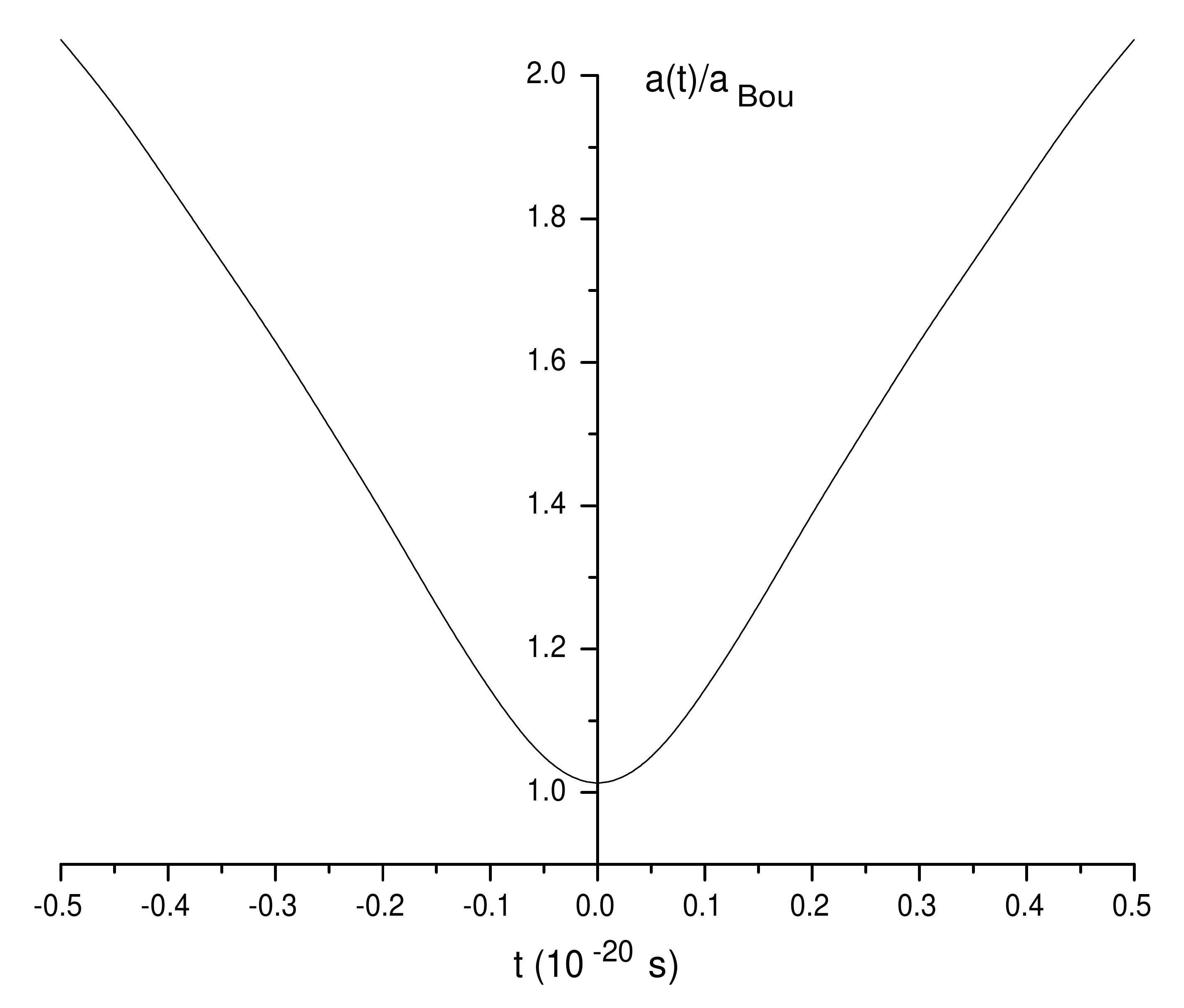}
\end{center}
\caption{Plot for the scale factor $a$ in function of time $t$. This figure
shows that the inflationary phase is very restrained.}
\label{figura4}
\end{figure}

The bouncing period plot is shown in figure \ref{figura4}.

\bigskip


Just looking for this plot it is possible to realize that the
primordial acceleration (inflation) engendered by the model is very
restricted. In fact, the equation (\ref{Fried_Bou4}) states that the
acceleration occurs only between $a_{Bou}<a(t)<1.32a_{Bou}$, which
in usual inflationary language corresponds just to $0.3$ e-folds. It
rules out the standard mechanism to solve de horizon problem.
Nevertheless, in this model the universe is cyclic and eternal, so
in fact this problem does not exist. Much more instigate (and
complicated) is the question of evolution of energy density
fluctuations. This issue is directly related with model stability
(see appendix B) and it will be subject of future investigations.

\subsection{The re-bouncing period and the recent acceleration}

\label{Sec-RBoun}

For the present, the Friedmann equations could be approximated by,
\begin{equation}
\left( \frac{\dot{a}}{a}\right) ^{2}\simeq H_{0}^{2}\left[ \Omega
_{m0}\left( \frac{a_{0}}{a}\right) ^{3}+\Omega _{\mu 0}\left( \frac{a}{a_{0}}%
\right) ^{4}+\Omega _{\beta 0}\left( \frac{a}{a_{0}}\right) ^{8}\right]
\label{Fried Rb1}
\end{equation}
and
\begin{equation}
\frac{\ddot{a}}{a}\simeq -\frac{H_{0}^{2}}{2}\left[ \Omega _{m0}\left( \frac{%
a_{0}}{a}\right) ^{3}-6\Omega _{\mu 0}\left( \frac{a}{a_{0}}\right)
^{4}-10\Omega _{\beta 0}\left( \frac{a}{a_{0}}\right) ^{8}\right]
\label{Fried Rb2}
\end{equation}
where $\Omega$'s are defined in (\ref{Omega_Definitions}).

The conditions for re-bouncing are given by
\[
\dot{a}_{RBou}=0 \text{ \, \, and \, \, } \ddot{a}_{RBou}<0.
\nonumber
\]

\begin{figure}[ht]
\begin{center}
\includegraphics[height=6.0cm, width=7.2cm]{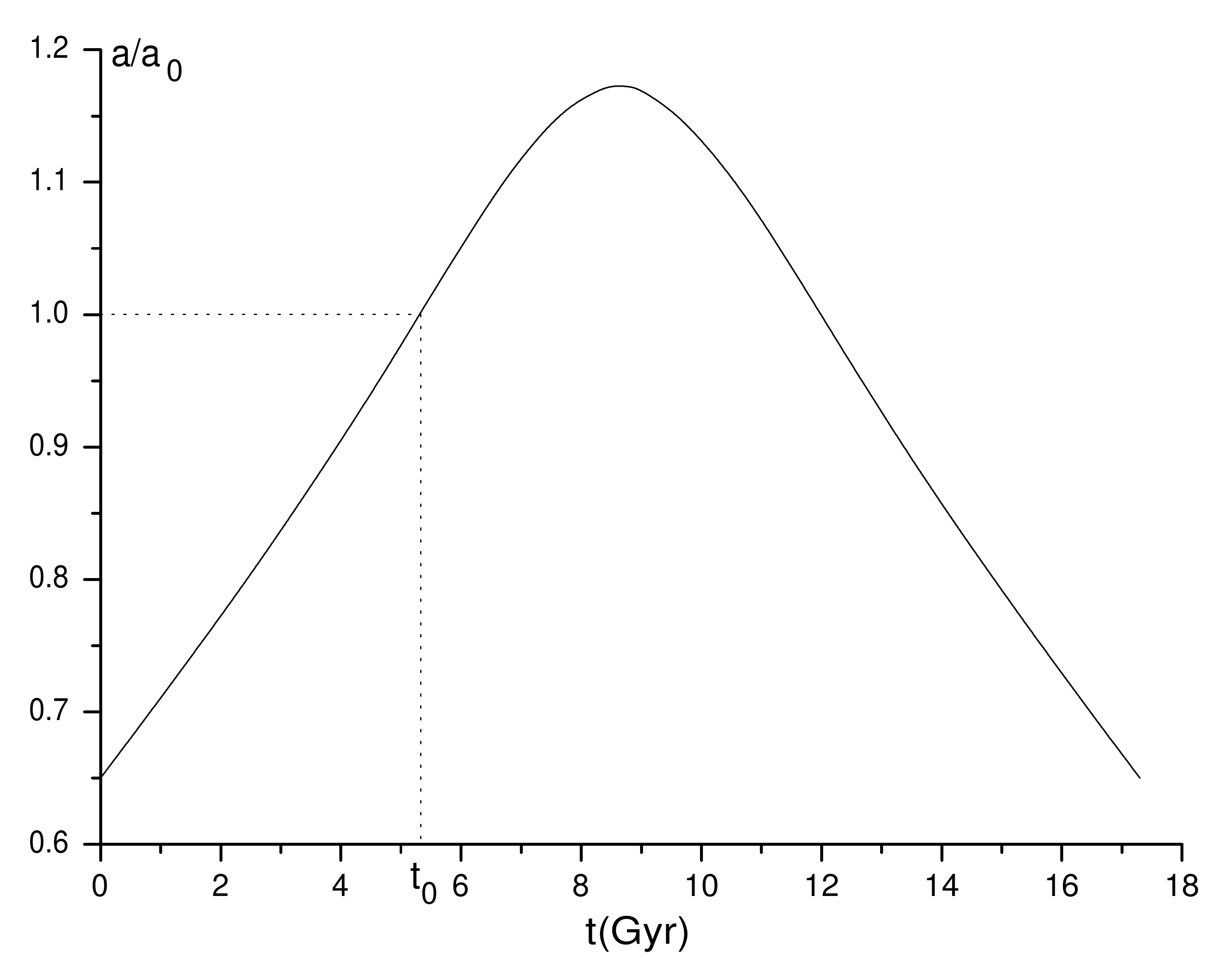}
\end{center}
\caption{Plot for the scale factor $a/a_{0}$ in function of time $t$ in Giga
years. $t=0$ is chosen as the instant when the actual acceleration begins. $%
t_{0}$ corresponds to interval of time between the start of acceleration and
the present time.}
\label{figura5}
\end{figure}

Unlike the equations of bounce, the equations above could not be
solved analytically. Even the $a_{RBou}$ could not be express from
$\Omega$'s in a simple manner. Nevertheless, numerical approaches
allow us to study many features of this period. Figure \ref{figura5}
shows the numerical solution for (\ref{Fried Rb1}).


From the numerical calculations involving (\ref{Fried Rb1}) it could be
determined the maximum size for the universe
\begin{equation}
a_{RBou}=1.173 a_{0}  \label{NumRboun}
\end{equation}
and the time remaining to reach it
\[
t_{RBou}-t_{0}=3.37 Gyr.
\]
The period of acceleration $\Delta t_{acel}$ is obtained from (\ref{Fried
Rb2}) and results in
\[
\Delta t_{acel}=5.33 Gyr.
\]
It started in $5.31$ billions of years ago and in terms of red-shift it
begun in
\[
z_{acel}=0.533.
\]

These results show that the acceleration era is almost finished, so spending
another 20 Million years we are entering in the re-bouncing period.

\subsection{The main features for the complete scenario}

To obtain global features for this model it is necessary to solve the first
Friedmann equation with all terms. It is given by

\begin{eqnarray}
\left( \frac{\dot{a}}{a}\right) ^{2} &=&H_{0}^{2}\left[ \Omega _{B0}\left(
\frac{a_{0}}{a}\right) ^{4}+\Omega _{UR0}\left( \frac{a_{0}}{a}\right)
^{4}-\Omega _{\alpha 0}\left( \frac{a_{0}}{a}\right) ^{8}\right. +  \nonumber
\\
&&+ \left. \Omega _{m0}\left( \frac{a_{0}}{a}\right) ^{3}+\Omega _{\mu
0}\left( \frac{a}{a_{0}}\right) ^{4}+\Omega _{\beta 0}\left( \frac{a}{a_{0}}%
\right) ^{8}\right]  \label{Fried_Comp}
\end{eqnarray}%
where $\Omega _{B0}\equiv \frac{B_{0}^{2}}{\rho _{c}}$ is always negligible.

\begin{figure}[ht]
\begin{center}
\includegraphics[height=6.0cm, width=7.2cm]{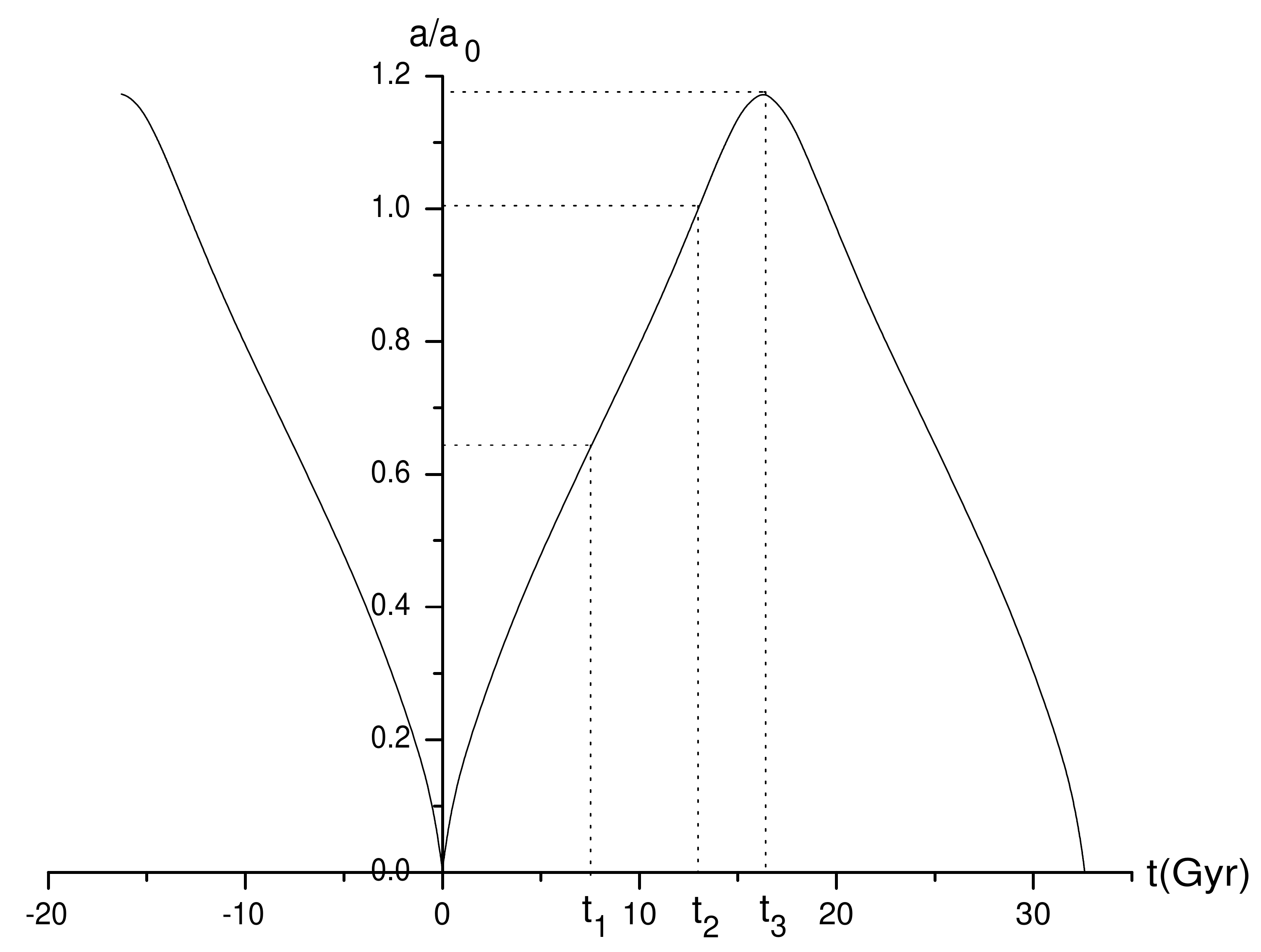}
\end{center}
\caption{Plot for the scale factor $a/a_{0}$ in function of time $t$ in Giga
years. $t=0$ is chosen as the instant of bounce. The three times $t_{1}$, $%
t_{2}$ and $t_{3}$ split the expansion period in three eras: matter era - $%
10^{-4}\lesssim t<t_{1}$, acceleration era - $t_{1}<t<t_{2}$ and re-bouncing
era $t_{2}<t<t_{3}$. The other two eras, radiation and bouncing eras, are
too close to origin, thus it is not possible distinguish them, at least in
theses scales.}
\label{figura6}
\end{figure}

Again we must use numerical methods to solve (\ref{Fried_Comp}). The result
is shown in figure \ref{figura6}.


The plot above shows the complete evolution for one and a half
cycle. This model is a eternal cosmological model, so it is
meaningless to talk about the initial time and the age of the
universe. Nevertheless, one could determine the time passed of
bounce until nowadays $\Delta t_{today}$ and
a period to complete a cycle $\Delta t_{cycle}$. Using the solution for (\ref%
{Fried_Comp}) it follows that $\Delta t_{today}=13.0 Gyr$ and $\Delta
t_{cycle}=32.6 Gyr$.

The solution of (\ref{Fried_Comp}) allows to establish the values
for the scale factor at the bouncing and re-bouncing instants. In terms of $%
a_{0}$ we obtain $a_{Bou}= 7.08\times 10^{-21} a_{0}$ and $a_{RBou}= 1.173
a_{0}$ where the maximum variation for the scale factor is $1.66\times
10^{20}$. Comparing these results with those obtained previously - equations
(\ref{Numboun}) and (\ref{NumRboun}) - it verifies that both have the same
numerical value. This confirms the validity of analysis made in sections \ref%
{Sec-Boun} and \ref{Sec-RBoun}.

Another interesting analysis which can be done is the evolution of
total equation of state $\omega_{T}$. Using the relation
\[
\omega_{T} = \frac{p_{T}}{\rho_{T}}
\]
we can study its evolution in terms of time. In most of the time
during a complete cycle $|\omega_{T}| \leq 1$. This range include
two long periods of deceleration (radiation and matter eras) and two
periods of acceleration (primordial and recent). However, because of
features of bouncing and re-bouncing, $\omega_{T}$ must eventually
assume values less than $-1$ and values greater than $1$. Indeed,
near to bouncing in a range $|t-t_{Bou}| < 1.1\times 10^{-21} s$ we
have $\omega_{T} < -1$, which means a phantom equation of state. And
near to re-bouncing in a range $|t-t_{RBou}| < 2.3 Gyr$ we have
$\omega_{T}
> 1$, i.e. an Ekpyrotic equation of state. It is noteworthy that
during all period of recent acceleration $\omega_{T} > -1$. This
analysis confirm that this model is an explicit realization of a
quintom cyclic scenario \cite{Quintoncyclic,QuintomCosmology}.

\section{Conclusion}

In this paper it was constructed a cyclic cosmological model with five fluids
namely baryonic and non-baryonic matter, radiation, neutrinos and a cosmic
magnetic field described by a nonlinear electrodynamics. With five
independent parameters it reproduces correctly the three expansion phases -
radiation, matter and accelerate phases - and produces a bouncing and a
re-bouncing stages. Another important point is that the main features for
the model could be analyzed splitting it in two independent phases
(primordial and actual) with a linking phase to connected both.

In primordial phase the relevant components are the bouncing and
ultra-relativistic components characterized by the constants $\alpha$ and $%
\Omega _{UR0}$. The ultra-relativistic constant was fixed through CMB
measurement and theoretical consideration about the neutrinos background. On
the other hand, the constant $\alpha$ remains widely free. It happens
because there are not much information about the pre-nucleosynthesis
universe. Another way to determine $\alpha$ is to interpret the term
$\alpha^{2} F^{2}$ as an one-loop quantum correction due vacuum
polarization in the infrared regime \cite{Euler,Schwinger}.
In this case, $\alpha^{2} \simeq 10^{-40}\rho_{c}^{-1}$ and this value is in
complete agreement with the cosmological constraints (see section \ref{Sec-CPU}).

Nowadays the relevant components are the non-relativistic matter, the
acceleration and the re-bouncing terms. The three constants - $\Omega _{m0}$%
, $\Omega _{\mu 0}$ and $\Omega _{\beta 0}$ - associated with each term was
fixed using data of SNIa. Figures \ref{figura1}, \ref{figura2} and \ref%
{figura3} show that the re-bounce probably occurs. For the complete
fit, represented by figure \ref{figura3}, the non-existence of
re-bouncing is excluded in $95\%$ confidence level. Other important
feature for this model
is the high value obtained for $\Omega _{m0}$ compared with the standard $%
\Lambda CDM$ model. Even with a robust prior for non-relativistic matter
this feature remains. I believe that it happens due the eminent deceleration
engendered by the re-bouncing term. To better clarify this point, it would
be important to use another cosmological tests. This possibility will be
investigated in the future.

A important question concerns about the evolution of energy density
fluctuations. As point out in section \ref{Sec-Boun}, the model not
engender an inflationary period. Nevertheless, it is not a essential problem
for eternal cyclic universe since the causal connection could be established
in previous cycle. Still, the subject about formation and/or dissociation of
structure (linked with the evolution of energy density fluctuations) is an
important topic and should be investigated in the future.

Other question concerns about the features of NLED used. As
mentioned in section \ref{Sec-MU}, the NLED was only used to
describe the cosmological magnetic field while the radiation was
retracted by the standard electrodynamics. So, the non-linear
effects act only in the magnetic field and the description of a free
radiation remains as in Maxwell theory. The main reason to adopt
this approach is because quantum corrections due vacuum polarization
changes the classical Lagrangian introducing nonlinear terms
\cite{Euler}. And the new terms only appear when we are treating
with a quasi-static electromagnetic field \cite{Schwinger}. So, the
fundamental Lagrangian is the Maxwell one and the extra terms are
just effects of vacuum polarization. This approach is directly
related with the model stability. Indeed, if we consider the
$L_{NLED}$ as fundamental Lagrangian instead of Maxwell Lagrangian
the model necessarily will present instabilities near to bounce (see
appendix B).

For a wide point of view, we can speculate that the electromagnetic
phenomenons in extremum conditions (e. g. very large and very short scale)
are described by a non-linear generalization of Maxwell electrodynamics.
Assuming that this generalization in a given scale could be expanded in
positive and negative powers of $F \equiv F_{\mu\nu} F^{\mu\nu}$, it is
necessary for convergence which $\alpha _{k}F^{k}>\alpha _{k+1}F^{k+1}$ and $%
\beta _{k}F^{-k}>\beta _{k+1}F^{-(k+1)}$ for $k>0$. So, it can be argued
that in this model the maximum and the minimum values for the scale factor
limit the influence of $F^{k}$ terms with $\left| k \right|>2 $, i.e. terms
of kind $F^{3}, F^{4}, ...$ and $F^{-3}, F^{-4}, ...$ are negligible in all
cosmological time. In this context, the proposal Lagrangian is an
approximation for a generic non-linear electrodynamics theory.

\section*{Acknowledgements}

I would like to acknowledge M. Novello, J. Salim and A. N. Ara\'{u}jo for
the very useful discussion and comments. I also acknowledge support of
FAPERJ.

\section*{Appendix A: Review of average procedure}

The standard cosmological model, provided by the FLRW geometry,
obeys the cosmological principle, which means that the 3-space is
homogeneous and isotropic geometry. Thus, for compatibility with the
cosmological framework, it is necessary an average procedure in the
electromagnetic field \cite{tolman}. As usual, we set the
volume-tric spatial average of a quantity $J$ by

\begin{equation}
\overline{J}\equiv \frac{1}{V}\int J\sqrt{-g}%
\;d^{3}x,
\end{equation}%
where $V=\int \sqrt{-g}\;d^{3}x$. And to generate the compatibility
with the FLRW geometry, it is necessary to impose that
\begin{equation}
\overline{E}_{i}=0,\;\;\;\overline{B}_{i}=0,\;\;\;\overline{E_{i}B_{j}}=0,
\end{equation}%
\begin{equation}
\overline{E_{i}E_{j}}=-\frac{1}{3}E^{2}g_{ij},\;\;\;\overline{B_{i}B_{j}}=-%
\frac{1}{3}B^{2}g_{ij}. \label{NL32}
\end{equation}%

Thus, applying the average procedure in the energy-momentum tensor
produced by a $L=L(F)$, we obtain a $T_{\mu \nu }$ of a perfect
fluid where
\begin{eqnarray}
\rho  &=&-L-4L_{\bar{F}}E^{2},  \nonumber \\
p &=&L-\frac{4}{3}\,(2B^{2}-E^{2})\,L_{\bar{F}},  \label{NL4}
\end{eqnarray}%
and $L_{\bar{F}}\equiv dL/d\bar{F}.$ The bar above $F$ point out
that spatial average was performed.

\section*{Appendix B: Model stability}

The purpose of this appendix is to discuss the model stability. As
it is well known all flat cosmological model with bounce violates
the null energy condition (NEC). On the other hand, in
\cite{Dubovsky} was shown that under fairly general conditions the
NEC violation necessarily implies instability (ghosts). More
specifically, the authors showed that when all null vectors violate
NEC (as in isotropic systems) or when superluminal excitations are
not present, the NEC violation will result in instabilities. These
statements extend in a straightforward manner for isotropic fluids
through the Lagrangian formulation for fluid dynamics \cite{fluid}.
Thus, it is reasonable to ask if the proposed model is stable?

To answer this question we will restrict the analysis to a region
near the bounce. In this case the relevant components are
represented by the following Lagrangian:

\begin{equation}
L\simeq\alpha^{2}F^{2}-\frac{1}{4}F-\frac{1}{4}F_{UR},\label{L_Boun}%
\end{equation}
where $F$ is associated with the magnetic field and $F_{UR}$ with
the ultrarelativistic constituents.


To analyze the stability of a model it is necessary to perturb this
model and analyze the behavior of these perturbations. Thus, in the
context of magnetic universe model, this analysis should be
separated in two cases: perturbations taken before the achievement
of spatial averages and perturbations taken after this achievement.

Let's discuss the first case (perturbations taken before). The term
$F^{2}$ arises when quantum corrections from vacuum polarization are
interpreted classically \cite{Euler,Schwinger}. Indeed, Schwinger
showed in \cite{Schwinger} that for quasi-static electromagnetic
fields the effects of vacuum polarization induce corrections in the
classical Lagrangian and the first of these corrections is given by
$\alpha^{2}F^{2}$. It is worth noting that the complete results
obtained in \cite{Euler} and \cite{Schwinger} are non-perturbative.
So, the term $\alpha^{2}F^{2}$ represents the inclusion (classical)
of effects of vacuum polarization in the cosmological background.
Therefore, the quantum perturbations should be done in the usual
Lagrangian of electrodynamics $\frac{1}{4}F$ which does not present
any instability problem. In this context, the quantization of
$L\simeq\alpha^{2}F^{2}-\frac{1}{4}F$ or any other $L_{NLED}$ is
completely meaningless.

For the second case, the perturbations must be performed in
energy-momentum tensor of perfect fluid (isotropic background). So,
the conclusions found in \cite{Dubovsky} can be directly applied,
and therefore the violation of NEC will imply in an unstable model.
Note that these perturbations should necessarily be of classical
type (such as gravitational perturbations) since the spatial average
has already been taken. In the magnetic universe model, this period
of instability occurs only near to bounce and it quickly ends (see
section \ref{Sec-Boun}).

From a different point of view, we can look at (\ref{L_Boun}) as a
fundamental Lagrangian. Thus, it is necessary to analyze the
stability of this Lagrangian. The term $-\frac{1}{4}F_{UR}$ will be
neglected because it does not generate instabilities. So,

\begin{equation}
L_{B}\simeq\alpha^{2}F^{2}-\frac{1}{4}F. \label{L_B}%
\end{equation}

The energy-momentum tensor of $L_{B}$ is given by

\[
T_{\mu\nu}=\left(  1-8\alpha^{2}F\right)  F_{\mu}^{\text{ \ }\alpha}%
F_{\alpha\nu}-\left(  \alpha^{2}F^{2}-\frac{1}{4}F\right)
g_{\mu\nu},
\]
remembering that $F_{\mu\nu}$ is determined only by magnetic field
$\vec{B}$.

To establish if there is a violation of NEC it is necessary analyze
the quantity $T_{\mu\nu}n^{\mu}n^{\nu}$ where $n^{\mu}=\left(
\omega,\vec {n}\right)  $ is a null vector. Performing this
calculation in FLRW background we obtain the following equation:
\begin{equation}
T_{\mu\nu}n^{\mu}n^{\nu}=\left[ 1-8\alpha^{2}F\right] \left[
\frac{\vec
{n}\times\vec{B}}{a}\right]  ^{2}, \label{NEC}%
\end{equation}
where $F$ is function of $\vec{B}^{2}$. Thus, to happen NEC
violation ($T_{\mu\nu}n^{\mu}n^{\nu}<0$) the first bracket should be
negative. Another important point is that if
$T_{\mu\nu}n^{\mu}n^{\nu}<0$ then \textit{all null vector} violate
NEC.

In general, the above equation is not spatially isotropic since
$\vec{B}$ may depend on points in space. Indeed, it is assumed that
$\vec{B}$ has a random spatial distribution whose spatial average is
determined by (\ref{NL32}). Therefore, it is possible there are
regions of space with instability (where NEC is violated for all
null vectors) and other regions perfectly stable. Note that the
instabilities only appear at many different points of space if, on
average, the term $8\alpha^{2}F$ is of the order (or larger) of
unity, i.e. $8\alpha^{2}\bar{F}$ $\gtrsim1$, and it is exactly what
happens near to bounce. So, using the bouncing condition
(\ref{boun}) and (\ref{L_B}) follows

\[
\rho_{Bou}\simeq\frac{1}{4}\bar{F}_{Bou}-\alpha^{2}\bar{F}_{Bou}%
^{2}=0\Rightarrow8\alpha^{2}\bar{F}_{Bou}=2,
\]
where $\bar{F}=2B^{2}$ - eq. (5).

Since, on average, the signal of (\ref{NEC}) is determined by%

\[
\left[  1-8\alpha^{2}\bar{F}\right]  \ \ \ \text{or \ }\ \left[
1-16\alpha^{2}B^{2}\right]  ,
\]
it is possible to conclude that near to bounce most regions will
present unstable perturbations. However, as the universe expands the
factor $B^{2}\sim a^{-4}$ decreases causing a decrease in the number
of regions with unstable perturbations. Note that with the increase
of $a$ in only one order of magnitude the probability of having a
region with unstable perturbations is negligible. One interesting
possibility for future work is to study how the perturbations
generated near to bounce evolve.


\begin{thebibliography}{99}
\bibitem{pbb} M. Gasperini and G. Veneziano, Phys. Rept. \textbf{373} 1
(2003), \texttt{hep-th/0207130}.

\bibitem{ekpy} J. Khoury, B. A. Ovrut, P. J. Steinhardt and N. Turok, Phys.
Rev. D \textbf{64}, 123522 (2001), \texttt{hep-th/0103239}.

\bibitem{Boun} M. Novello and S.E. Perez Bergliaffa, Phys. Rep. \textbf{463}%
, 127 (2008).

\bibitem{leo} R. Aldrovandi, R. R. Cuzinatto and L. G. Medeiros, Eur. Phys.
J. C \textbf{58}, 483 (2008), \texttt{arXiv:0801.0705}.

\bibitem{Herman} C. Corda and H. J. M. Cuesta, Astropart. Phys. \textbf{34},
587 (2011), \texttt{arXiv:1011.4801}.

\bibitem{nelson} P. Peter and N. Pinto-Neto, Phys. Rev. D \textbf{78},
063506 (2008).

\bibitem{novello1} V. De Lorenci, R. Klippert, M. Novello and J. M. Salim,
Phys. Rev. D \textbf{65}, 063501 (2002).

\bibitem{novello2} M. Novello, S.E. Perez Bergliaffa and J. Salim, Phys.
Rev. D \textbf{69}, 127301 (2004), \texttt{astro-ph/0312093}.

\bibitem{port1} P. Vargas Moniz, Phys. Rev. D \textbf{66}, 103501
(2002).

\bibitem{port2} V. Dyadichev, D. Gal'tsov, A. Zorin and M. Zotov, Phys. Rev. D
\textbf{65}, 084007 (2002), \texttt{hep-th/0111099}.

\bibitem{port3} R. Garcia-Salcedo and N. Breton, Int. J. Mod. Phys. A \textbf{%
15}, 4341 (2000), \texttt{gr-qc/0004017}.

\bibitem{port4}  M. Sami, N. Dadhich and Tetsuya Shiromizu, Phys. Lett. B
\textbf{568}, 118 (2003), \texttt{hep-th/0304187}.

\bibitem{port5} E. Elizalde, J. Lidsey, S. Nojiri and S. Odintsov, Phys. Lett. B \textbf{574}%
, 1 (2003), \texttt{hep-th/0307177}.

\bibitem{port6} R. Garcia-Salcedo, N. Breton, Class. Quant. Grav. \textbf{20}, 5425 (2003),
\texttt{hep-th/0212130}.

\bibitem{Aline} M. Novello, A. N. Araujo and J. Salim, Int. J. Mod. Phys. A
\textbf{24}, 5639 (2009), \texttt{arXiv:0802.1875}.

\bibitem{Riess1998} A. G. Riess et al., Astron. J. \textbf{116}, 1009
(1998), \texttt{astro-ph/9805201}.

\bibitem{Perm} S. Permutter et al., Nature \textbf{391}, 51 (1998), \texttt{%
astro-ph/9712212}.

\bibitem{Quintoncyclic} H. H. Xiong, Y. F. Cai, T. Qiu, Y.S. Piao and X.
Zhang, Phys. Lett. B \textbf{666}, 212 (2008),
\texttt{arXiv:0805.0413}.

\bibitem{cyclic1} J. V. Narlikar, G. Burbidge and R.G. Vishwakarma,
J. Astrophys. Astron. \textbf{28}, 67 (2007),
\texttt{arXiv:0801.2965}.

\bibitem{QuintonExt1} Y. F. Cai and E. N. Saridakis, JCAP \textbf{10}, 20
(2009), \texttt{arXiv:0906.1789}.

\bibitem{QuintonExt2} Y. F. Cai and E. N. Saridakis, Class. Quant. Grav.
\textbf{28}, 35010 (2011), \texttt{arXiv:1007.3204}.

\bibitem{QuintonExt3} Y. F. Cai and E. N. Saridakis, J. of Cosm. \textbf{17},
7238 (2011), \texttt{arXiv:1108.6052}.

\bibitem{Euler} W. Heisenberg and H. Euler, Z. Phys. \textbf{98}, 714 (1936).

\bibitem{Schwinger} J. Schwinger, Phys. Rev. \textbf{82}, 664 (1951).

\bibitem{PDG} Particle Data Group, site - http://pdg.lbl.gov/ (2008).

\bibitem{WeinCosm} S. Weinberg, \textit{Cosmology}, Oxford University Press,
New York 2008.

\bibitem{Mag} S. Bertone et al., MNRAS 370, 319 (2006), \texttt{%
astro-ph/0604462}; F. Govone and L. Feretti, Int. J. Mod. Phys. D \textbf{13}%
, 1549 (2004), \texttt{astro-ph/0410182}.

\bibitem{Union} M. Kowalski et al., ApJ \textbf{686}, 749 (2008), \texttt{%
arXiv:0804.4142}.

\bibitem{Sas} M. Sasaki, MNRAS \textbf{228}, 653 (1987).

\bibitem{Hol_Lin} D. E. Holz and E. V. Linder, ApJ \textbf{631}, 678 (2005).

\bibitem{Astier} P. Astier et al., A\&A \textbf{447}, 31 (2006), \texttt{%
astro-ph/0510447}.

\bibitem{Schindler} S. Shindler, Space Science Reviews \textbf{100}, 299
(2002), \texttt{astro-ph/0107028}.

\bibitem{Allen} S. W. Allen et al., MNRAS \textbf{383}, 879 (2008), \texttt{%
arXiv:0706.0033}.

\bibitem{Kirkman} D. Kirkman et al., ApJ Suppl. \textbf{149}, 1 (2003),
\texttt{astro-ph/0302006}.

\bibitem{Fredmann} W. L. Freedman et al., ApJ \textbf{553}, 47 (2001),
\texttt{astro-ph/0012376}.

\bibitem{QuintomCosmology} Y. F. Cai, E. N. Saridakis, M. R. Setare and J. Q.
Xia, Phys. Rep. \textbf{493}, 1 (2010), \texttt{arXiv:0909.2776}.

\bibitem{tolman} R. Tolman and P. Ehrenfest, Phys. Rev. \textbf{36}, 1791
(1930).

\bibitem{Dubovsky} S. Dubovsky et al., JHEP \textbf{603}, 25 (2006), \texttt{%
hep-th/0512260v2}.

\bibitem{fluid} R. Jackiw et al., J. Phys. A \textbf{37}, 327
(2004), \texttt{hep-ph/0407101}.

\end{thebibliography}
\end{document}